\documentclass[twocolumn,prb]{revtex4}
\usepackage{amsfonts}
\usepackage[T1]{fontenc}
\usepackage{amsmath,amsbsy,amssymb,graphicx}
\usepackage{times}

\let\mathbf=\boldsymbol

\begin{document}

\title{Higher-order topological electric circuits and topological corner
resonance\\
on the breathing Kagome and pyrochlore lattices}
\author{Motohiko Ezawa}
\affiliation{Department of Applied Physics, University of Tokyo, Hongo 7-3-1, 113-8656,
Japan}

\begin{abstract}
Electric circuits are known to realize topological quadrupole insulators. We
explore electric circuits made of capacitors and inductors forming the
breathing Kagome and pyrochlore lattices. They are known to possess three
phases (trivial insulator, higher-order topological insulator and metallic
phases) in the tight-binding model, where the topological phase is characterized
by the emergence of zero-energy corner states. A topological phase
transition is induced by tuning continuously the capacitance, which is
possible by using variable capacitors. It is found that the two-point
impedance yields huge resonance peaks when one node is taken at a corner in
the topological phase. It is a good signal to detect a topological phase
transition. We also show that the topological corner resonance is robust
against randomness of capacitance and inductance. Furthermore, the size of
electric circuits can be quite small to realize the topological phase
together with topological phase transitions.
\end{abstract}

\maketitle

\textit{Introduction:} Topological insulators and its generalization to
higher-order topological insulators \cite%
{Fan,Science,APS,Peng,Lang,Song,Bena,Schin,FuRot,EzawaKagome,Khalaf, Switch}
are fascinating topics in condensed-matter physics (CMP). They are
characterized by the bulk symmetry and the bulk topological numbers, and
observed by the emergence of topological zero-energy boundary states.
Especially, topological zero-energy corner states emerge for the
second-order topological insulators (SOTI) in two dimensions and for the
third-order topological insulators in three dimensions. They are robust
against impurities. They are studied mainly in fermionic systems in materials%
\cite{Bis,EzawaPhos,MoTe}. Actually, it is quite difficult to make
experimental observation of topological corner states in CMP. Furthermore,
although topological phase transitions have been extensively studied, the
experimental observation is also very difficult in CMP. On the other hand,
these topological corner states have already been observed experimentally in
phononic system\cite{Phonon,Xue,Alex}, microwave system\cite{Microwave},
photonic system\cite{Photon} and electric circuits\cite{TECNature}.

Topological corner states in square lattice are experimentally realized in
electric circuits\cite{TECNature,Garcia}. Indeed, the Su-Schrieffer-Heeger
model\cite{ComPhys,Hel}, the honeycomb lattice\cite{ComPhys,Hel} and Weyl
semimetals\cite{ComPhys,Lu} have already been implemented in electric
circuits. The impedance is a measurable quantity determining whether the
system is topological or not, where topological boundary resonance effects
occur in the topological phases. Here we note that the emergence of
topological corner states has been predicted\cite{EzawaKagome} also in the
breathing Kagome and pyrochlore lattices in the context of CMP. Thus, it is
an interesting problem to study measurable quantities in topological
electric circuits corresponding to these lattices.

\begin{figure}[t]
\centerline{\includegraphics[width=0.49\textwidth]{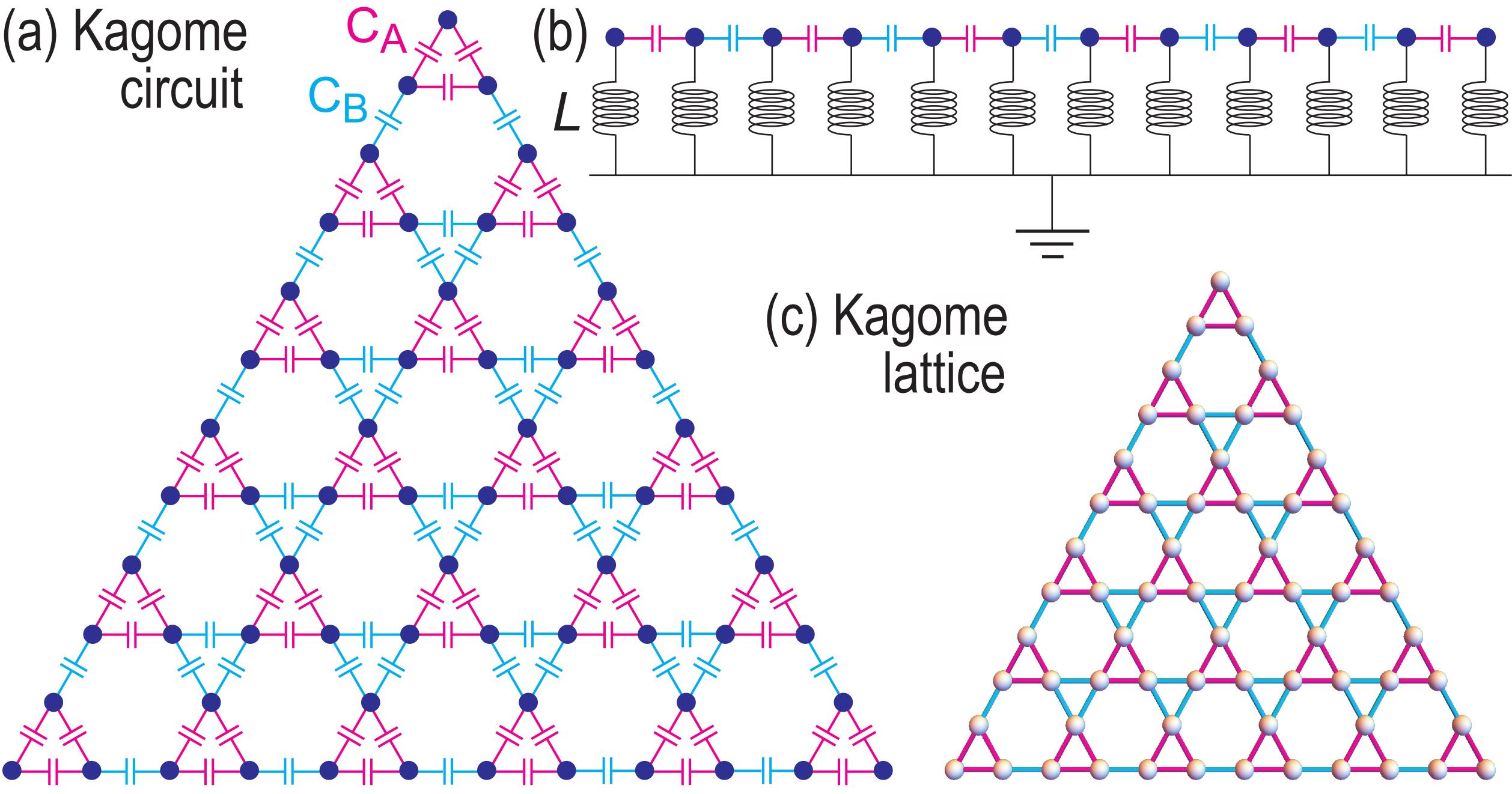}}
\caption{Illustration of the breathing Kagome circuit composed of two types
of capacitors (with capacitance $C_{A}$ and $C_{B}$) and inductors (with
inductance $L$). Adjacent nodes are connected by capacitors and each node is
grounded by an inductor. The size of the triangle is $\ell =6$.}
\label{FigKagomeCircuit}
\end{figure}

Let us explain how to construct a topological electric circuit by taking an
instance of the breathing Kagome lattice. The breathing Kagome lattice
consists of lattice sites and two types of links indicated in red and cyan
as in Fig.\ref{FigKagomeCircuit}(c). We insert capacitors with capacitance 
$C_{A}$ and $C_{B}$ to links in red and cyan, respectively, as in Fig.\ref{FigKagomeCircuit}(a). 
Then, we connect each lattice site to the ground via
an inductor with inductance $L$, as illustrated in Fig.\ref{FigKagomeCircuit}(b). 
A lattice site is called a node in electric circuit. It is clear that
this method is applicable to any lattices we encounter in CMP.

In this paper, we study electric circuits corresponding to the breathing Kagome
and pyrochlore lattices. Topological phase transitions in electric circuits
are well signaled by measuring the impedance, where huge resonance peaks
emerge at corners in the topological phase. We find the topological
robustness, that is, this resonance is robust against randomness of
capacitance and inductance. We explicitly investigate a triangular geometry
made of the breathing Kagome circuit, where we define its size $\ell $ by
the number of small upper triangles along one edge: See Fig.\ref{FigKagomeCircuit}. 
We also study a tetrahedron geometry made of the breathing pyrochlore circuit.

\textit{Topological electric circuits:} Electric circuits are characterized
by the Kirchhoff's current law\cite{TECNature,ComPhys,Hel}, 
\begin{equation}
\frac{d}{dt}I_{a}=\sum_{b}C_{ab}\frac{d^{2}}{dt^{2}}\left(
V_{a}-V_{b}\right) +\frac{1}{L_{a}}V_{a},
\end{equation}
where $I_{a}$ is the current between node $a$ and the ground, $V_{a}$ is the
voltage at node $a$, $C_{ab}$ is the capacitance between nodes $a$ and $b$, 
$1/L_{a}$ is the inverse of the inductance at node $a$, and the sum is taken over all adjacent nodes $b$. 
See an example of the breathing Kagome circuit in Fig.\ref{FigKagomeCircuit}.
When we apply an AC field $V\left( t\right)
=V\left( 0\right) e^{i\omega t}$, the Kirchhoff's law is rewritten as 
\begin{equation}
I_{a}\left( \omega \right) =\sum_{b}J_{ab}\left( \omega \right) V_{b}\left(
\omega \right)
\end{equation}
with 
\begin{equation}
J_{ab}\left( \omega \right) =i\omega \left[ C_{ab}+\delta _{ab}\left(
\sum_{c}C_{ac}-\frac{1}{\omega ^{2}L_{a}}\right) \right] ,  \label{EqJ3}
\end{equation}
where the matrix $\mathbf{J}\left( \omega \right) =\{J_{ab}\left( \omega
\right) \}$ is called the circuit Laplacian. It is a linear operator and
corresponds to a tight-binding Hamiltonian $H$ in CMP
via the relation $J_{ab}\left( \omega \right) =i\omega H_{ab}\left( \omega
\right) $ with the Hamiltonian being\cite{TECNature,ComPhys} 
\begin{equation}
H_{ab}\left( \omega \right) =C_{ab}+\delta _{ab}\left( \sum_{c}C_{ac}-\frac{1}{\omega ^{2}L_{a}}\right) .
\end{equation}
The capacitor between adjacent nodes $a$ and $b$ corresponds to the
transfer integral $t_{ab}\leftrightarrow C_{ab}$ between adjacent sites $a$
and $b$, while the inductor attached to node $a$ corresponds to the on-site
potential $U_{a}\leftrightarrow \sum_{b}C_{ab}-(1/\omega ^{2}L_{a})$ at the
site $a$. Later we present an explicit correspondence in the case of the
breathing Kagome lattice.

By diagonalizing the matrix $\mathbf{J}\left( \omega \right) $ we obtain the
eigenvalue $j_{n}$ and the associated eigenmode $\left\vert \mathbf{\psi }_{n}\right\rangle $. 
Then, we have $\mathbf{J}\left( \omega \right)
=\sum_{n}j_{n}\left\vert \mathbf{\psi }_{n}\right\rangle \left\langle 
\mathbf{\psi }_{n}\right\vert $. The eigenmode $\left\vert \mathbf{\psi }_{n}\right\rangle $ 
is a vector whose components are labelled by node $a$; 
$\left\vert \mathbf{\psi }_{n}\right\rangle =\{\psi _{n,a}\}$. The admittance
eigenvalue $j_{n}$ is a measurable quantity\cite{Hel}.

The two-point impedance is given by\cite{ComPhys,TECNature} 
\begin{equation}
Z_{ab}=\frac{V_{a}-V_{b}}{I_{ab}}=\sum_{n}\frac{\left\vert \psi _{n,a}-\psi
_{n,b}\right\vert ^{2}}{j_{n}},  \label{EqImpB}
\end{equation}
and determined by measuring the voltage response by running a current
between two nodes $a$ and $b$. The key property is that $Z_{ab}$ diverges in
the presence of zero-admittance modes ($j_{n}=0$) provided $\psi _{n,a}\neq
\psi _{n,b}$. Hence, the emergence of zero-admittance modes may be detected
by measuring the two-point impedance.

\begin{figure}[t]
\centerline{\includegraphics[width=0.49\textwidth]{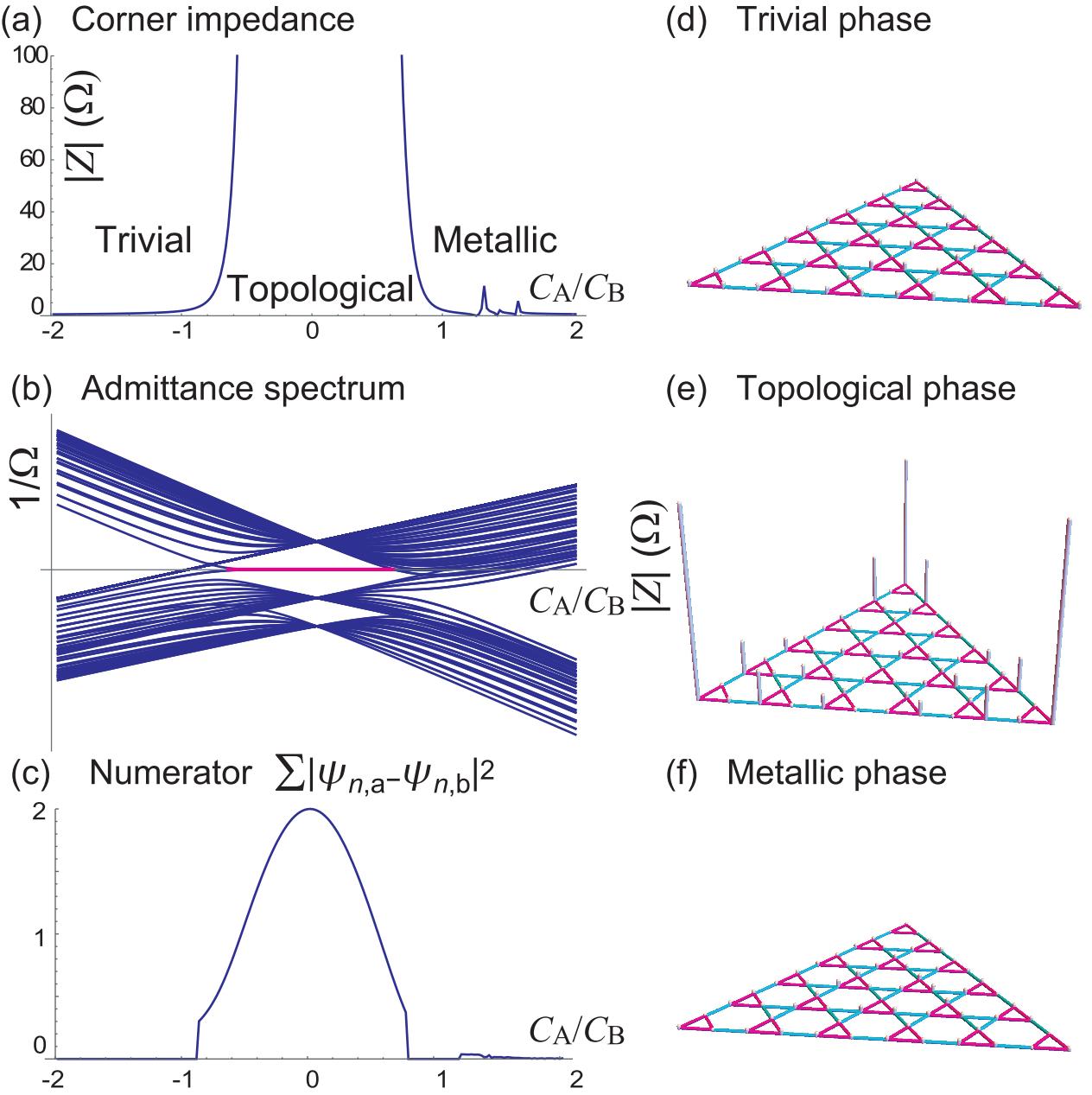}}
\caption{Two-point impedance for the breathing Kagome circuit. (a) The
maximum value of the two-point impedance as a function of $C_{A}/C_{B}$. (b)
The admittance spectrum at the resonant frequency $\protect\omega =\protect\omega _{\text{c}}$ 
as a function of $C_{A}/C_{B}$. (c) The numerator of the
impedance $\sum_{n=1}^{3}|\protect\psi _{n,a}-\protect\psi _{n,b}|^{2}$.
Spatial distribution of two-point impedance (d) in the trivial phase, (e) in
the topological phase, and (f) in the metallic phase. One node is fixed in
the vicinity of the triangle center. Absolute value of the impedance is
represented by the length of the tubes.
We have taken $C_{B}=1\protect\mu $F and $L=1\protect\mu $H. We use
a triangle with $\ell =6$.}
\label{FigKagomeBand}
\end{figure}

\begin{figure*}[t]
\centerline{\includegraphics[width=0.96\textwidth]{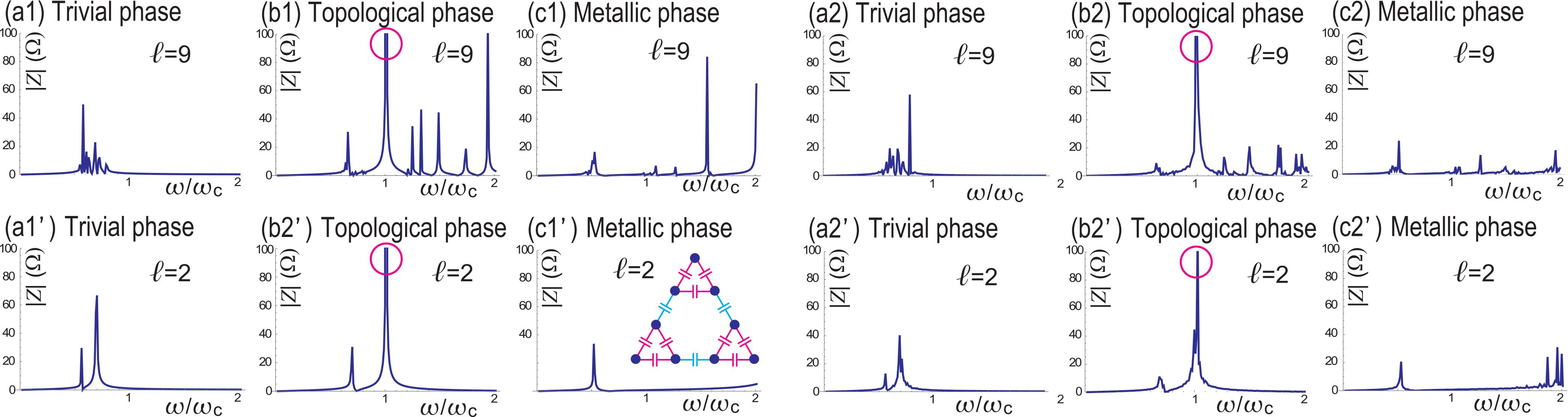}}
\caption{Corner impedance $|Z|$ as a function of $\protect\omega /\protect\omega _{\text{c}}$ 
of the triangle made of the breathing Kagome circuit 
(a) in the trivial phase ($C_{A}/C_{B}=-1.5$), (b) in the topological phase ($C_{A}/C_{B}=0.1$) 
and (c) in the metallic phase ($C_{A}/C_{B}=1.5$). Red circles
indicate the resonance peak arizing from topological corner modes. The
height of the peak is as huge as $10^{9}$ $\Omega$ in the case of $\ell=9 $.
The size $\ell$ is shown in the figure. (a1)--(c1') Corner impedance $|Z|$
without randomness; (a2)--(c2') Corresponding impedance $|Z|$ in the
presence of $5\%$ randomness.
The inset of (c1') illustrates the breathing Kagome circuit with the size $\ell=2$.
}
\label{FigCorner}
\end{figure*}

\textit{Breathing Kagome circuit:} The electric circuits corresponding to the
honeycomb lattice have already been studied\cite{ComPhys,Hel}. Here we
investigate them for the breathing Kagome lattice, which is known to realize
a SOTI in CMP. We consider an infinite circuit which is periodic with a unit
cell. It corresponds to a bulk system in CMP. 

The circuit Laplacian (\ref{EqJ3}) for an infinite circuit reads 
\begin{equation}
J=i\omega \left[ 2\left( C_{A}+C_{B}\right) -\frac{1}{\omega ^{2}L}\right] 
\mathbb{I}-i\omega H_{\text{Kagome}},  \label{EqJ2}
\end{equation}
where $\mathbb{I}$ is the unit matrix and 
\begin{equation}
H_{\text{Kagome}}=\left( 
\begin{array}{ccc}
0 & h_{12} & h_{13} \\ 
h_{12}^{\ast } & 0 & h_{23} \\ 
h_{13}^{\ast } & h_{23}^{\ast } & 0
\end{array}
\right) ,  \label{EqJ1}
\end{equation}
with 
\begin{eqnarray}
h_{12} &=&C_{A}+C_{B}e^{-i\left( k_{x}/2+\sqrt{3}k_{y}/2\right) },  \notag \\
h_{13} &=&C_{A}+C_{B}e^{-ik_{x}},  \notag \\
h_{23} &=&C_{A}+C_{B}e^{i\left( -k_{x}/2+\sqrt{3}k_{y}/2\right) }.
\end{eqnarray}
Here, $C_{A}$ and $C_{B}$ are capacitances shown in Fig.\ref{FigKagomeCircuit}(a). 
We note that the Hamiltonian $H_{\text{Kagome}}$ is
precisely the same one that describes the tight-binding model for the breathing
Kagome lattice by replacing $C_{A}$ and $C_{B}$ with the hopping
parameters $t_{a}$ and $t_{b}$, respectively: See Eq.(1) of Ref.\cite{EzawaKagome}. 
Consequently, the system (\ref{EqJ1}) for the breathing Kagome circuit is
topological for $-1<C_{A}/C_{B}<1/2$, trivial for $C_{A}/C_{B}<-1$ and
metallic for $C_{A}/C_{B}>1/2$. Consequently, the system undergoes
topological phase transitions at $C_{A}/C_{B}=-1$ between the trivial and
topological phases, and at $C_{A}/C_{B}=1/2$ between the topological and
metallic phases. In contrast to the case of CMP, it will be rather easy to
make experimental observation of these phase transitions by tuning the
capacitance continuously. We note that negative capacitance is possible\cite{TECNature} 
with the use of inductors by identifying $C\equiv -1/\omega ^{2}L
$.

We investigate the topological phase in triangular geometry [Fig.\ref{FigKagomeCircuit}(a)], 
where topological zero-admittance modes are present
at the corners. Due to the presence of zero-admittance modes, the second
term in the right-hand side of Eq.(\ref{EqJ2}) vanishes. The resultant
equation is a standard formula for the LC circuit with capacitance 
$C_{A}+C_{B}$. The resonant frequency is given by the zero of the identity
matrix and given by $\omega _{\text{c}}=1/\sqrt{2L\left( C_{A}+C_{B}\right) }$. 

The behavior of the impedance around $\omega _{\text{c}}$ is expressed as 
\begin{equation}
\left\vert Z\right\vert \propto 1/\left( \omega ^{2}-\omega _{\text{c}}^{2}\right) ,  \label{EqImpA}
\end{equation}
which yields a huge resonance peak at the frequency $\omega _{\text{c}}$. 
There is no divergence because of the finite-size effect.
On the other hand, when there are no zero-admittance modes, the impedance is
finite. The metallic phase is intriguing due to the presence of the sea of
zero-admittance modes. As we shall see soon, there is no resonance
enhancement in $\left\vert Z\right\vert $. We expect that the emergence of
the resonant modes is a signal that the electric circuit is in a topological
phase.

When we use the capacitor of the order of 1$\mu $F and the inductor of
the order 1$\mu $H, the resonance occurs around 1MHz and the impedance is of
the order of 1$\Omega $, while the resonant impedance becomes to the order of $10^9\Omega$.

\textit{Corner impedance:} We consider a triangle structure made of the
breathing Kagome circuit [Fig.\ref{FigKagomeCircuit}]. We first show the
admittance spectrum in Fig.\ref{FigKagomeBand}(b), where zero-admittance
corner modes emerge only in the topological phase. 

We next investigate the two-point impedance. We fix one node $a$\
arbitrarily, and measure the impedance $Z_{ab}$ between node $a$ and
another node $b$. By moving $b$ over all nodes, we obtain a space
distribution of the two-point impedance. We show the results in the three
phases in Fig.\ref{FigKagomeBand}(d)--(f), where node $a$ is taken around
the center of the triangle. The essential feature is a strong enhancement of
the two-point impedance in the topological phase when node $b$ is taken at
three corners. We have found that this essential feature does not depend on
the position of the fixed node $a$ provided it is not taken on the corners.
When node $a$ is taken on a corner, the strong enhancement appears only when node $b$ is taken at
the other two corners because $Z_{aa}=0$. The huge peak in $Z_{ab}$ is
easily understood in the topological phase due to zero-admittance corner
modes as we have discussed below Eq.(\ref{EqImpA}). 
We find that the strongest resonance occurs when two nodes $a$ and $b$ are taken at two
different corners.

We show the two-point impedance in Fig.\ref{FigCorner}, where the two nodes
are fixed at two different corners. We show the impedance as a function of 
$\omega /\omega _{\text{c}}$. The impedance displays a huge peak at $\omega =\omega _{\text{c}}$ 
in the topological phase, while there are no such peaks in the
trivial phase and the metallic phase. We also show the impedance at the
resonant frequency $\omega _{\text{c}}$ as a function of $C_{A}/C_{B}$ in
Fig.\ref{FigKagomeBand}(a). It becomes huge rapidly in the topological phase,
which implies that it is a good indicator to observe topological phases. 
Remarkably, the resonance peak signaling the topological phase
is clearly present in such a small triangle that has the size $\ell =2$: See Fig.\ref{FigCorner}(a')--(c').

Naively, we expect that the impedance takes a large value also for metallic
phase since there are many zero-admittance modes although they are not
topological. However, this is not the case. We show the numerator 
$\left\vert \psi _{n,a}-\psi _{n,b}\right\vert ^{2}$ as a function of $C_{A}/C_{B}$, 
where the sum of $n$ is taken only for the three
zero-admittance modes in Fig.\ref{FigKagomeBand}(c). It takes value around 2
only for the topological phase representing the two localization of the
corner modes. On the other hand, in the metallic phase, it is very small, 
$\left\vert \psi _{n,a}-\psi _{n,b}\right\vert ^{2}\propto 1/N$, where $N$ is
the number of nodes. Accordingly, the impedance is small in the metallic
phase although there are plenty of zero-admittance modes.

\textit{Effects of randomness:} We next study the effects of randomness in
capacitors and inductors. For this purpose, we make substitution 
$C_{i}\mapsto C_{i}\left( 1+\eta _{i}\right) $ and $L_{i}\mapsto L_{i}\left(
1+\xi _{i}\right) $, where $\eta _{i}$ and $\xi _{i}$ are uniformly
distributed random variables ranging from $-\delta $ to $\delta $. We have
calculated the impedance by choosing $\delta =0.05$.

We show the $\omega $ dependence of the impedance in Fig.\ref{FigCorner}.
The prominent peak signaling the topological resonance remains as it is. On
the other hand, all other peaks are reduced. The results indicate the
topological robustness of the topological corner resonance.

\begin{figure}[t]
\centerline{\includegraphics[width=0.49\textwidth]{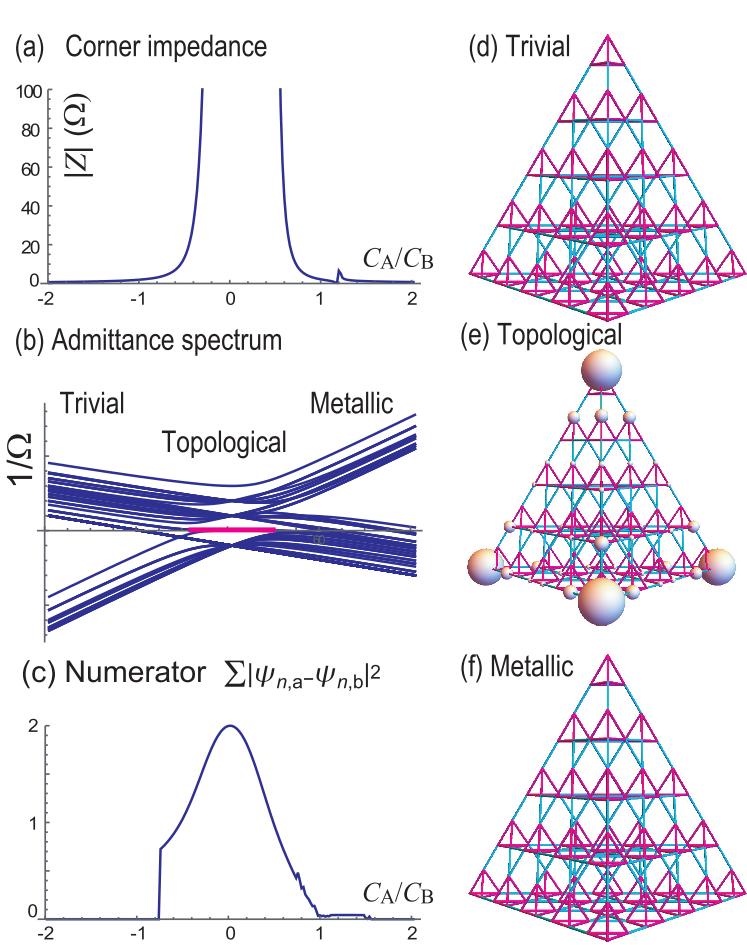}}
\caption{Two-point impedance for the breathing pyrochlore circuit. (a) The
maximum value of the two-point impedance as a function of $C_{A}/C_{B}$. (b)
The admittance structure at the resonant frequency $\protect\omega =\protect\omega _{\text{c}}$ 
as a function of $C_{A}/C_{B}$. (c) The numerator of the
impedance $\sum_{n=1}^{4}|\protect\psi _{n,a}-\protect\psi _{n,b}|^{2}$.
Spatial distribution of two-point impedance (d) in the trivial phase, (e) in
the topological phase, and (f) in the metallic phase. One node is fixed in
the vicinity of the tetrahedron center. Absolute value of the impedance is
represented by the size of a ball. Huge balls are found at four corners in
the topological phase. We have taken $C_{B}=1\protect\mu $F and $L=1\protect\mu $H.}
\label{FigPyro}
\end{figure}

\textit{Breathing pyrochlore circuit:} A natural extension of the breathing
Kagome circuit to three dimensions is the breathing pyrochlore circuit,
where a third-order topological insulator is realized\cite{EzawaKagome}. The
circuit Laplacian is given by 
\begin{equation}
L=i\omega \left[ 3\left( C_{A}+C_{B}\right) -\frac{1}{\omega ^{2}L}\right] 
\mathbb{I}-i\omega H_{\text{pyro}},
\end{equation}
where 
\begin{equation}
H_{\text{pyro}}=\left( 
\begin{array}{cccc}
0 & h_{12} & h_{13} & h_{14} \\ 
h_{12}^{\ast } & 0 & h_{23} & h_{24} \\ 
h_{13}^{\ast } & h_{23}^{\ast } & 0 & h_{34} \\ 
h_{14}^{\ast } & h_{24}^{\ast } & h_{34}^{\ast } & 0
\end{array}
\right) ,
\end{equation}
with 
\begin{eqnarray}
h_{12} &=&C_{A}+C_{B}e^{-i\left( k_{x}+k_{y}\right) /2},  \notag \\
h_{13} &=&C_{A}+C_{B}e^{-i\left( k_{y}+k_{z}\right) /2},  \notag \\
h_{14} &=&C_{A}+C_{B}e^{-i\left( k_{z}+k_{x}\right) /2},  \notag \\
h_{23} &=&C_{A}+C_{B}e^{-i\left( k_{z}-k_{x}\right) /2},  \notag \\
h_{24} &=&C_{A}+C_{B}e^{-i\left( -k_{y}+k_{z}\right) /2},  \notag \\
h_{34} &=&C_{A}+C_{B}e^{-i\left( k_{x}-k_{y}\right) /2}.
\end{eqnarray}
The resonant frequency is $\omega _{\text{c}}=1/\sqrt{3L\left(
C_{A}+C_{B}\right) }$. Topological phase diagram of the breathing pyrochlore
circuit is the same as that of the breathing Kagome circuit. We show the
admittance spectrum of the tetrahedron in Fig.\ref{FigPyro}(b), where the
four topological corner modes appear in the topological phase. We show the
two-point impedance between two nodes as a function of $C_{A}/C_{B}$ in Fig.\ref{FigPyro}(a), 
which becomes huge in the topological phase. We also show the
numerator $\left\vert \psi _{n,a}-\psi _{n,b}\right\vert ^{2}$, where the
sum of $n$ is taken only for the four zero-admittance modes in Fig.\ref{FigPyro}(c).
A space distribution of the two-point impedance is shown in the three phases in Fig.\ref{FigPyro}(d)--(f). 

\textit{Discussion:} We have shown that the topological corner impedance is
a good signal to detect a topological phase transition 
in electric circuits corresponding to the breathing Kagome and pyrochlore lattices, 
where the huge resonance peak emerges only in the topological phase. The topological phase
transition is controlled by tuning variable capacitors. It is not necessary
to tune the capacitance so precisely because of the topological robustness.
Furthermore, to realize the topological phase together with topological
phase transitions, the size of the electric circuit can be quite small.

The author is very much grateful to N. Nagaosa for helpful discussions on
the subject. This work is supported by the Grants-in-Aid for Scientific
Research from MEXT KAKENHI (Grants No. JP17K05490, No. JP15H05854 and No.
JP18H03676). This work is also supported by CREST, JST (JPMJCR16F1).

\end{document}